# Using effective medium theories to design tailored nanocomposite materials for optical systems

Daniel Werdehausen*[a,b,c], Isabelle Staude[b], Sven Burger[d,e], Jörg Petschulat[a,f], Toralf Scharf[c], Thomas Pertsch[b,g], and Manuel Decker[a]

[a]Corporate Research & Technology, Carl Zeiss AG, Carl Zeiss Promenade 10, 07745 Jena, Germany; [b]Institute of Applied Physics, Abbe Center of Photonics, Friedrich Schiller University Jena, Albert-Einstein-Str. 15, 07745 Jena, Germany; [c]Nanophotonics and Metrology Laboratory, École Polytechnique Fédérale de Lausanne (EPFL), CH-1015 Lausanne, Switzerland; [d]JCMwave GmbH, Bolivaralle 22, 14050 Berlin, Germany; [e]Zuse Institute Berlin, Takustr. 7, 14195 Berlin, Germany; [f]Semiconductor Mask Solutions, Carl Zeiss SMT GmbH, Carl Zeiss Promenade 10, 07745 Jena, Germany; [g]Fraunhofer Institute of Applied Optics and Precision Engineering, Albert-Einstein-Str. 7, 07745 Jena, Germany

*Daniel.Werdehausen@zeiss.com;

## ABSTRACT

Modern optical systems are subject to very restrictive performance, size and cost requirements. Especially in portable systems size often is the most important factor, which necessitates elaborate designs to achieve the desired specifications. However, current designs already operate very close to the physical limits and further progress is difficult to achieve by changing only the complexity of the design. Another way of improving the performance is to tailor the optical properties of materials specifically to the application at hand. A class of novel, customizable materials that enables the tailoring of the optical properties, and promises to overcome many of the intrinsic disadvantages of polymers, are nanocomposites. However, despite considerable past research efforts, these types of materials are largely underutilized in optical systems. To shed light into this issue we, in this paper, discuss how nanocomposites can be modeled using effective medium theories. In the second part, we then investigate the fundamental requirements that have to be fulfilled to make nanocomposites suitable for optical applications, and show that it is indeed possible to fabricate such a material using existing methods. Furthermore, we show how nanocomposites can be used to tailor the refractive index and dispersion properties towards specific applications.

**Keywords:** Nanocomposites, Nanomaterials, Optical materials, Optical design, Optical Systems, Dispersion engineering, Effective medium theory, Mie theory

## 1. INTRODUCTION

Nanocomposites promise to be a class of materials with tunable and novel properties, and have therefore been at the center of considerable research efforts both in optics [1-11] and other fields [12-15]. In optics, most of the attention has been directed towards high-refractive index coatings, which can for example be used to enhance the efficiencies of LEDs [11].

However, nanocomposites are also an attractive class of materials for more complicated optical systems, e.g. imaging systems that contain a high number of optical elements and materials. The reason for this is that the optical properties of nanocomposites can be tailored by changing not only the constituent materials, but also the concentration, size, and shape of the nanoparticles. This would provide a significant benefit for optical systems, since the interplay of different aberrations critically depends on the material choice. In addition, constraints on the cost and manufacturability often further limit the available selection of materials. The choice and optimization of the materials is therefore one of the most important and difficult steps in optical design, and, in practice, relies on the optical designer's experience, trial and error, and discrete optimization.

The quantities that are commonly used in optical design to characterize the optical properties of a material are the refractive index $n_d$, where the subscript refers to the d-line ($\lambda_d = 587.56$ nm), the Abbe number ($\nu_d$) and partial dispersion ($P_{g,F}$), which are defined as [16]:

$$\nu_d = \frac{n_d - 1}{n_F - n_C}, \quad \text{and} \quad P_{g,F} = \frac{n_g - n_F}{n_F - n_C}, \quad (1)$$

where the subscripts again refer to the Fraunhofer lines ($\lambda_g = 435.83$ nm, $\lambda_F = 435.13$ nm, and $\lambda_C = 656.27$ nm). The definitions in equation (1) show that the Abbe number is a general measure of the material's overall dispersion in the visible regime, whereas the partial dispersion quantifies the amount of dispersion at wavelengths on the blue side of the spectrum. The Abbe number consequently characterizes the amount of chromatic aberration a single optical element introduces, whereas the partial dispersion is needed to quantify the residual chromatic aberrations of more complicated optical systems [16]. Other examples of how the material choice affects the amount of aberrations are that the amount of spherical aberration is reduced, if a material with a higher refractive index is used, or that the Petzval field curvature can be minimized by combining materials with high and low refractive indices [16]. However, in reality, only a specific region of combinations of $n_d$, $\nu_d$, and $P_{g,F}$ is accessible, since there is a fundamental connection between the magnitude of the refractive index and the amount of dispersion [17]. A problem, which is even exacerbated if the application or fabrication technique relies on a limited choice of materials, as is, for example, the case in 3D printing [18-20]. In this paper we therefore investigate, if nanocomposites can be used to overcome some of these limitations, and provide a platform with tunable optical properties.

This paper is organized as follows: to show how customized nanocomposites can be designed and optimized, we will first give a summary on effective medium theories that are suitable for the modelling of nanocomposites. We give a thorough introduction into this area, since a plethora of different effective medium theories has been published, and it can often be confusing to choose the correct effective medium theory, and determine the conditions under which it is valid. Furthermore, a look at the derivation of the effective medium theories also provides valuable information, as to how nanocomposites have to be designed to make them suitable for optics. In the second part, we will then show that due to the intrinsic heterogeneity of nanocomposites, different effective medium regimes have to be distinguished, and discuss that optical nanocomposites can only be realized in a specific parameter range. Finally, we will briefly investigate the potential of such materials for optical system design.

## 2. MODELLING OF NANOCOMPOSITES

### 2.1 Effective mediums and the Clausius-Mossoti equation

An effective medium is a material that consists of two or more distinguishable components, but whose optical properties can nevertheless be described by an (effective) refractive index. Effective medium theories have been developed to model the properties of such materials. For the design of tailored composite materials, these theories are indispensable tools that complement conventional optical design software packages, which are based on ray optics, and can therefore only be used to optimize the material's refractive index, Abbe number, and partial dispersion. Since such a treatment doesn't account for wave optical effects, effective medium theories have to be used to determine how a nanocomposite has to be designed in terms of the materials, shapes, concentrations, and particle sizes so that it also fulfills all requirements. Over the last few decades a multitude of different effective medium theories has been developed [21-35]. These theories, in general, can be classified into two categories [25]: Firstly, the ones that deal with so-called "separated-grain structures", in which a high number of inclusions are randomly distributed in a host medium, and the host and the inclusions can consequently be clearly distinguished. These types of mediums can be modeled using Maxwell-Garnett-type theories [25]. In contrast, in the second case, the "aggregate structures", a clear identification of one material as the host and another material as the inclusions is not possible, and the two materials randomly occupy certain volumes in space. Materials in this regime can be treated using Bruggeman's model [25]. In this paper, we focus on the former case, which is applicable when nanoparticles are dispersed in a host material. But our analysis can be straightforwardly extended to aggregate-type materials.

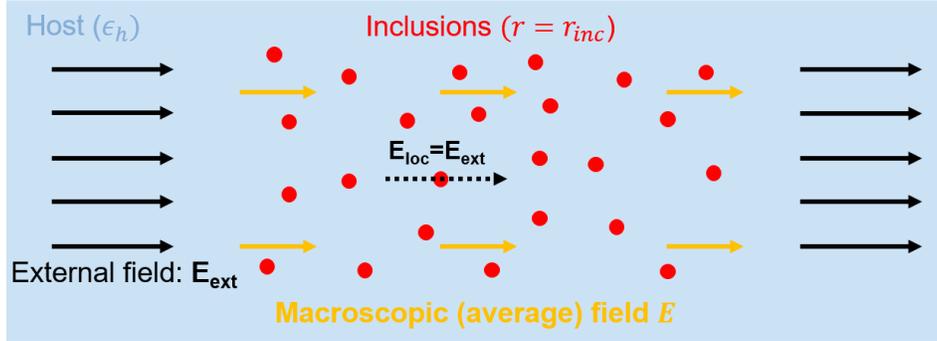

Figure 1. Schematic of an effective medium with a separated-grain structure, in which distinct particles are randomly dispersed in a homogenous host with permittivity $\epsilon_h$. The local electric field ($\mathbf{E_{loc}}$), which is identical to the external field ($\mathbf{E_{ext}}$) for perfectly randomly distributed nanoparticles, polarizes the inclusions. The macroscopic field (**E**) is smaller than $\mathbf{E_{ext}}$ due to the dipole fields from the inclusions.

To discuss the potential and limitations of Maxwell-Garnett-type effective medium theories, it is important to understand the approximations on which they are based. To this end, we will briefly discuss their derivation. The general geometry we consider for this purpose is depicted in fig. 1: Inclusions, which we will later require to be spherical, are embedded in a host. We assume the host to form a homogenous background, in that its optical properties can be described by a dielectric function ($\epsilon_h(\lambda)$). If now an external electric field $\mathbf{E_{ext}}$ is applied, it induces an electromagnetic response in the inclusions that, in general, can be written as a multipole expansion including all electric and magnetic multipoles. However, if we require the particles to be very small compared to the wavelength, the higher order multipoles can be neglected, and only the electric dipole term has to be considered [23]. This restriction will be lifted to some extend later on, where we also include the magnetic dipole term. In the electric dipole limit, the total polarization ($\mathbf{p_{tot}}$) of the inclusions in a unit volume can consequently be written as $\mathbf{p_{tot}} = \sum_{i=1}^{N} \mathbf{p}_i = N\langle \mathbf{p} \rangle = N\alpha\langle \mathbf{E_{loc}} \rangle$, where $\mathbf{p}_i$ is the dipole moment of the i-th dipole, N is the number density (particle concentration) of the inclusions, $\langle \mathbf{p} \rangle$ is the average dipole moment per inclusion, α is the dipole polarizability, and $\langle \mathbf{E_{loc}} \rangle$ is the average local field at the positions of the inclusions. However, within a macroscopic volume, the fields of the inclusions oppose the external field and the macroscopic field **E** in the composite medium is therefore smaller than $\mathbf{E_{ext}}$. An expression for this macroscopic field **E**, which is the field that has to be substituted into the macroscopic Maxwell's equations in case of a composite medium, can be obtained by averaging the local field over a volume that is small compared to the wavelength, but contains a high number of inclusions. The result of this averaging procedure reads:

$$\mathbf{E} = \left(1 - \frac{8\alpha f}{\varepsilon_h d_{inc}^3}\right) \langle \mathbf{E}_{loc} \rangle, \qquad (2)$$

where we introduced the volume fraction $f = 4/3\, N\pi(d_{inc}/2)^3$, and assumed the particles to be spherical (diameter $d_{inc}$). If we now assume a perfectly random distribution, it can be shown that the contributions to the local field at the position of one particular particle from all other particles cancel out, so that the average local field is identical to the external field ($\langle \mathbf{E_{loc}} \rangle = \langle \mathbf{E_{ext}} \rangle$)[32]. This also holds true for particles arranged in a simple cubic lattice, but it is straightforward to see that this assumption breaks down if there is a correlation between the positions of the inclusions, for example due to agglomeration. The fact that $\langle \mathbf{E_{loc}} \rangle = \langle \mathbf{E_{ext}} \rangle$ is valid for a perfectly random distribution is also the reason why the effective medium theories remain accurate in the case of multiple scattering up to surprisingly large volume fractions. However, there are certain limits to the theories' accuracy at very high volume fractions, since impenetrable nanoparticles with a finite volume can never be perfectly dispersed (f > 20 %) [33].

To derive an expression for the effective permittivity of the composite material, the microscopic description outlined so far has to be combined with the macroscopic expression for the total polarization of a finite volume V. To this end, the total polarization can be written as $\boldsymbol{p_{tot}} = V[(\epsilon_{eff} - \epsilon_h)/4\pi]\boldsymbol{E}$, where $\epsilon_{eff}$ is the effective permittivity of the composite medium. By combining this expression with its microscopic equivalent and making use of equation (2), we now arrive at the Clausius-Mossoti (CM) equation for inclusions in a homogenous host:

$$\frac{\varepsilon_{eff} - \varepsilon_h}{\varepsilon_{eff} + 2\varepsilon_h} = \frac{8f}{d_{inc}^3}\alpha, \qquad (3)$$

which connects the (microscopic) polarizability of the inclusions to the (macroscopic) effective permittivity of the composite material. This expression itself is already an effective medium theory, and is the foundation of all Maxwell-Garnett-type theories.

## 2.2 The limits of the concept of an "effective optical medium"

The above derivation highlights that, in general, any material is made up of a multitude of scatters, and it would therefore be necessary to deal with a multibody scattering problem. However, it can be shown rigorously that this multibody problem, can macroscopically be reduced to a scalar refractive index if the scatterers in the material act as ideal dipoles, and form a homogenous distribution on the scale of the wavelength [29, 36]. Microscopically this can be explained by the fact that the radiation from individual scatters interferes in such a way that a plane wave in such a material simply propagates at a lower speed than in vacuum, while its shape remains unaffected [36]. This allows to determine the macroscopic electromagnetic fields both inside and outside of the material for any source field using the well known macroscopic boundary conditions [37].

To overcome the limitation to dipolar scatterers for the description of nanocomposites, one might be tempted to include higher orders in the multipole expansion. However, one should keep in mind that this complicates the propagation of electromagnetic waves in the medium. This can be seen if one considers that the electric displacement (**D**) for a medium with a quadrupole component in *(ω,r)*-space reads [38-40]:

$$\boldsymbol{D}(\omega, \boldsymbol{r}) = \epsilon_0 \boldsymbol{E}(\omega, \boldsymbol{r}) + \boldsymbol{P}(\omega, \boldsymbol{r}) - \frac{1}{2}\nabla \boldsymbol{Q}(\omega, \boldsymbol{r}), \qquad (4)$$

where $\boldsymbol{Q}$ is the quadrupole tensor, which itself is a function of $\nabla\boldsymbol{E}$ [40, 41]. In Fourier-space these spatial derivatives can be replaced by a simple scalar product with the wave vector $\boldsymbol{k}$, which makes it possible to define a permittivity [39, 40]:

$$\boldsymbol{D}(\boldsymbol{k}, \omega) = \epsilon_0 \epsilon(\boldsymbol{k}, \omega) \boldsymbol{E}(\boldsymbol{k}, \omega). \qquad (5)$$

This shows, that a medium with a quadrupole component is characterized by a $\boldsymbol{k}$-dependent permittivity $\epsilon(\boldsymbol{k}, \omega)$, and hence also a $\boldsymbol{k}$-dependent refractive index $n(\boldsymbol{k}, \omega)$. This, in turn, leads to spatial dispersion (non-local effects). Moreover, the validity of equations (4) and (5) again relies on the scatterers forming a homogenous distribution. However, this assumption is clearly violated for actual nanoparticles that exhibit a significant quadrupole component, since they are no longer much smaller than the wavelength, and the number of nanoparticles per volume $\lambda^3$ consequently fluctuates heavily around the mean. This leads to incoherent scattering [23]. This is why such a medium should be considered as "heterogeneous" [42]. This demonstrates that composite mediums in general should only be referred to as a real "effective medium" if the material fulfils both the homogeneity, and the dipole requirements.

## 2.3 The original Maxwell-Garnett-theory

The final step in the derivation of the original Maxwell-Garnett (MG) theory is to obtain an expression for the polarizability (α) in terms of the constituents' permittivities. In the point dipole limit, this can be done in the so-called quasistatic approximation, in which the field is assumed to be constant across the particle. In this limit, the polarizability can be determined by solving Laplace's equation both inside and outside the particle, and enforcing the appropriate boundary conditions [32]. The results reads:

$$\alpha_{stat} = (d_{inc}/2)^3 \frac{\epsilon_{inc} - \epsilon_h}{\epsilon_{inc} + 2\epsilon_h}. \qquad (6)$$

This expression, however, it is only valid for particles for which retardation, i.e. the phase difference across the particle, is negligible, and which consequently act as point dipoles.

## 2.4 Going beyond the quasistatic limit

To lift the restriction to point dipole scatterers, and thus include particles with larger dimensions, it is necessary to leave the quasistatic regime. To this end, an extension to original MG theory is required. For spherical particles this can be done using Mie theory, which gives the rigorous solution for the electromagnetic fields in and around a sphere. Mie theory, in general, expresses the electromagnetic response of a sphere as a multipole expansion [23]. Since Mie theory is a rigorous solution, this allows to determine the full dipole polarizability outside of the quasistatic limit directly from the dipole term of the expansion [22]:

$$\alpha_{\text{Mie}} = i\frac{3(d_{\text{inc}}/2)^3}{2x^3}a_1, \qquad (7)$$

where $a_1$ is the first order Mie coefficient, which is a function of the size parameter $x = \pi n_{\text{h}} d_{\text{inc}}/\lambda$, and the relative refractive index $= n_{\text{inc}}/n_{\text{h}} = \sqrt{\epsilon_{\text{inc}}}/\sqrt{\epsilon_{\text{h}}}$. By substituting this expression into the CM equation (equation (3)) we obtain the so-called Maxwell-Garnett-Mie theory. Note, that by expanding equation (7) into orders of the size parameter, and neglecting all but the first term, one arrives back at the quasistatic expression (equation (6)). This illustrates that the higher order terms in the expansion account for retardation within the dipole term.

So far we've only considered the electric dipole terms, and discussed that if higher order electric multipoles contribute to the nanoparticles' response the medium is no longer an "effective optical medium". The magnetic dipole, however, can be easily incorporated into the effective medium framework, by considering that the presence of magnetic dipole radiation leads to non-unity permeability:

$$\mu_{\text{eff}} = \frac{x^3 + 3ifb_1}{x^3 - \frac{3}{2}ifb_1}. \qquad (9)$$

The refractive index then readily follows from $\boldsymbol{n} = \sqrt{\epsilon_{\text{eff}}\mu_{\text{eff}}}$. This, in principle, would allow to design an "effective magnetic material", even if all constituents are non-magnetic [42]. Microscopically the existence of the magnetic component is caused by the presence of circular displacement currents around the nanoparticles [43]. Note, however, that in order for the light to couple to the magnetic dipole mode, a certain amount of retardation is necessary.

## 2.5 Limitations of Maxwell-Garnett-type theories for small particles

Both the quasistatic and the Mie expressions for the polarizability are only valid, as long as the optical properties of the inclusions can be described by a permittivity. This limits the applicability of both theories in the small particle limit, since the permittivity is a macroscopic quantity that is obtained by averaging over a volume that contains a large number of atoms/molecules, yet is small compared to the wavelength [36]. Furthermore, it has to be kept in mind that in the single-digit nanometer range, where the nanoparticles still consist of a sufficiently large number of atoms for the permittivity to still be well defined, the permittivity depends on the particle size. In fact, it is well known from both experimental and theoretical studies that for dielectric particles quantum confinement starts to affect the permittivity at radii in the order of the exciton Bohr radius [44]. This, in general, leads to a widening of the electronic gap [45], and therefore to a reduction of the refractive index (for photon energies below the band gap). In contrast to purely dielectric particles, the permittivity of metallic nanoparticles, also depends dominantly on the contribution from free electrons. Due to surface scattering, this contribution also exhibits a size-dependence, which has to be considered for particle sizes in the order of the conduction electrons' mean free path. This leads to an increased amount of electron scattering, which can be accounted for by introducing a size dependent damping rate [46]:

$$\epsilon_{\text{inc}}(\omega, r_{\text{inc}}) = \epsilon_{\text{bulk}} + \frac{\omega_{\text{p}}^2}{\omega^2 + i\omega\gamma_0} - \frac{\omega_{\text{p}}^2}{\omega^2 + i\omega\left(\gamma_0 + \frac{2Av_{\text{f}}}{d_{\text{inc}}}\right)}, \qquad (10)$$

where $\omega_{\text{p}}$ and $\gamma_0$ are the plasma frequency, and the damping constant of the bulk material respectively, $v_{\text{f}}$ is the Fermi velocity, and $A$ is a proportionality factor. The second term in this expression corresponds to the Drude permittivity for free electrons with a reversed sign. This provides an intuitive interpretation of equation (10), namely that the expression replaces the contribution from the free electrons in the bulk material, by a free electron contribution whose damping rate is increased due to surface scattering.

# 3. EFFECTIVE MEDIUM REGIMES

## 3.1 Regime classification

As already discussed, the concept of a conventional refractive index is based on the assumptions that the individual scatters are dipoles, and form a homogenous distribution on the scale of the wavelength. A violation of these requirements, in case of composite materials, manifests itself macroscopically in the onset of incoherent scattering. In this section we will discuss that this fundamental connection between particle size and scattering necessitates the distinction of different effective medium regimes, and limits the validity of the concept of an effective refractive index for composite materials containing larger inclusions.

In general, the use of a refractive index implies that plane waves retain their shape as they propagate through the material (see sec. 2.2) and are only attenuated because of absorption. For conventional, homogenous optical materials, the imaginary part of the refractive index can hence be used to predict both the transmission through, as well as the absorption/heating rate within the medium. However, this only remains true for composite materials, if the inclusions are small enough for scattering to be negligible, which is only the case if the fundamental assumptions of homogeneity and dipolarity are not violated. If these conditions are fulfilled, the material is an "unrestricted effective mediums", in that the effective refractive index of such a material can be used with the same validity as the one of a conventional optical material (table 1) [42]. In this regime, the effective refractive index consequently still allows determining the macroscopic fields both inside and outside of the material using Snell's law and the Fresnel equations.

In contrast, if scattering plays a role, absorption will no longer be the only mechanism that removes energy from a beam of light, and it is straightforward to see that an effective refractive index can no longer be used to determine the distribution of the electromagnetic fields. Such an effective refractive index therefore provides no information about the appearance of the material: a high imaginary part could imply that the material appears black due to absorption, or milky because of scattering. However, if scattering represents only a small slow perturbation, and the beam retains its directionality over a certain distance, it can still be useful to define an "effective refractive index" to describe the behavior of the coherent part of the beam [24]. In this regime, the real part of the effective refractive index can still be used in Snell's law to determine beam's direction, whereas the imaginary part only provides information about how much energy remains within the beam and not about the microscopic origin of the losses. This is why such a material is a "restricted effective medium", in the sense that the effective refractive index only has a restricted validity (table 1) [42].

Finally, if there is a large amount of scattering, and a beam of light loses its directional character within short propagation distances, the material enters the "heterogenous regime" (table 1). In this regime, the concept of an effective refractive index no longer as any validity (sec. 2.2). As already discussed this regime is reached for nanocomposites if high order multipoles contribute to the nanoparticles' response. Furthermore, in practice, the regime of a certain nanocomposite is not only determined by the particle size, but also by the blending of the nanoparticles into the host matrix. If the nanoparticles are not perfectly integrated into the molecular structure of the host, the homogeneity requirement will also be violated, and large amount of scattering will be the consequence.

Table 1. Different effective medium regimes. The regime a nanocomposite is located in mainly depends on the particle size. But agglomeration or a non-ideal blending of the nanoparticles into the host can also lead to scattering, and therefore move a material into the restricted or heterogeneous regimes.

| Unrestricted effective medium regime | Restricted effective medium regime | Heterogeneous regime |
|---|---|---|
| • No scattering<br>• $n_{eff}$ can be used without limitations | • Scattering plays a role<br>• $n_{eff}$ only describes behavior of coherent part of beam → Can't be used to predict appearance | • High amount of scattering<br>• The concept of a refractive index is no longer valid<br>• Higher order multipoles contribute for nanocomposites |

## 3.2 Requirements on optical nanocomposites and regime boundaries

To ensure a high image quality, optical imaging systems require materials that are free of scattering, since stray light significantly affects image quality. To make nanocomposites suitable for such applications, it is therefore critical that they are designed and fabricated as "unrestricted effective mediums". To elucidate at what particle size scattering starts to play a role, we, in this section, analyze the transition from the unrestricted to the restricted effective medium regime, using the effective medium theories introduced in sec. 2.

As already discussed, the original Maxwell-Garnett effective medium theory operates in the quasistatic dipole limit, and therefore does not account for scattering losses (equation (6)). The Maxwell-Garnett-Mie theory, in contrast, can predict scattering losses to a certain degree, since it is based on the full dipole polarizability (equation (7)). To investigate at what particle size scattering starts to affect the material properties, fig. 2 depicts the transmission of a beam of light through a 1 mm thick slab of a $ZrO_2$ nanocomposite at a volume fraction of 20% and different particle sizes. We chose $ZrO_2$ because it is a widely investigated system that can be mass-produced [9, 10]. The transmission curves in fig. 2 comprise both the results from the original Maxwell-Garnett theory (dotted line), and the Maxwell-Garnett-Mie theory with the magnetic dipole term (solid lines). To ensure comparability, we did not account for the size dependence of the permittivity. The figure shows that the original Maxwell-Garnett theory predicts the losses to be negligible, which illustrates that absorption does not play a role in this system, since this theory does not account for scattering losses. However, the fact that the results from the Maxwell-Garnett-Mie theory deviate from that the ones from the original theory even at particle sizes as small as 2 nm, illustrates that the quasistatic approximation is only valid for particle sizes below 2 nm. Furthermore, the drastic dependence of the transmission on the particle size demonstrates that scattering starts to severely affect the system's properties even at particle sizes as small as 4 nm. This shows that high-quality nanocomposites, which can be used in optical elements with a thickness of several millimeters or more, have to contain nanoparticles which are (significantly) smaller than 4 nm. This particle size consequently defines the upper boundary of the "unrestricted effective medium regime".

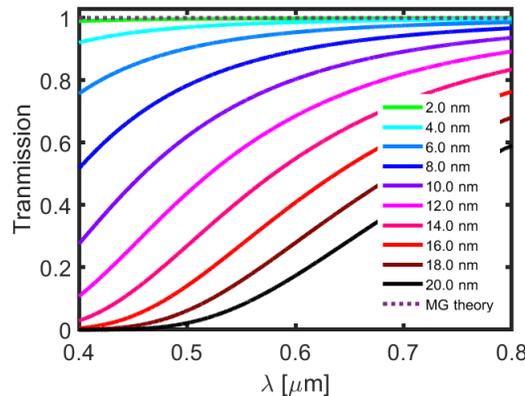

Figure 2. Transmission through a 1 mm thick slab of a $ZrO_2$-nanocomposite in PMMA. The dotted line corresponds to the results from the original Maxwell-Garnett theory, whereas the results at different particle sizes were obtained from the Maxwell-Garnett-Mie theory with the dipole magnetic dipole contribution. It can be seen that scattering causes the transmission to decrease at particle sizes around 4nm.

## 4. BENEFITS OF OPTICAL NANOCOMPOSITES

The existence of the "unrestricted effective medium regime", in which scattering does not play a role proofs that it is possible to design nanocomposites that are suitable for imaging applications. To evaluate the potential of this approach for optical system design, we dedicate this section to a brief discussion of its benefits for optical systems.

To illustrate how nanocomposites can be used to tune the optical properties, Fig. 3 depicts $n_d$ and $v_d$ as a function of the volume fraction for the $ZrO_2$ nanocomposite investigated in the previous section. The figure demonstrates that adding the nanoparticles into the polymer significantly changes both quantities, and allows adjusting them continuously along a trajectory that is defined by the constituent materials. Furthermore, the refractive indices that are achieved at high concentrations are significantly higher the ones of conventional polymers. Especially the combination of a high refractive index, and a relatively large Abbe number cannot be achieved with conventional polymers [47]. Since, in principle, different nanoparticle materials can be combined with different hosts, developing a material platform comprising different nanoparticle materials and hosts would consequently allow accessing new refractive index regimes, and tuning the optical properties continuously over a relatively broad range. Within this range, it would be possible to treat the optical properties as free parameters in optical design, and optimize them for the application at hand.

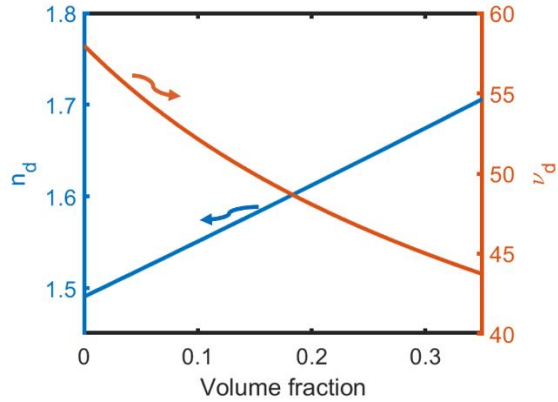

Figure 3. $n_d$ and $v_d$ as a function of the volume fraction for a $ZrO_2$-PMMA nanocomposite at $d_{inc}$ = 4 mm. For a fixed material combination, varying the concentration allows to tune the optical properties along a fixed trajectory.

To demonstrate the potential of this approach on the example of a simple prototype system, fig. 4 depicts the designs of a $f = 4$ mm lens with an aperture size of $d = 4$ mm both with and without a nanocomposite material. For the reference design without nanoparticles, we used normal PMMA, whereas we again chose the $ZrO_2$/PMMA nanocomposite at a volume fraction of $f = 35\%$ for the design with the nancomposite material. The optical properties of both materials can be directly obtained from fig. 3 at volume fractions of 0% and 35%, respectively. The comparison of the designs illustrate that adding the nanoparticles leads to a significant volume reduction (28%), and a drastic increase of the performance. The reason for the improved performance is the reduced amount of spherical aberration, which manifests itself in a reduction of the spot radius from 278 µm to 124 µm (RMS).

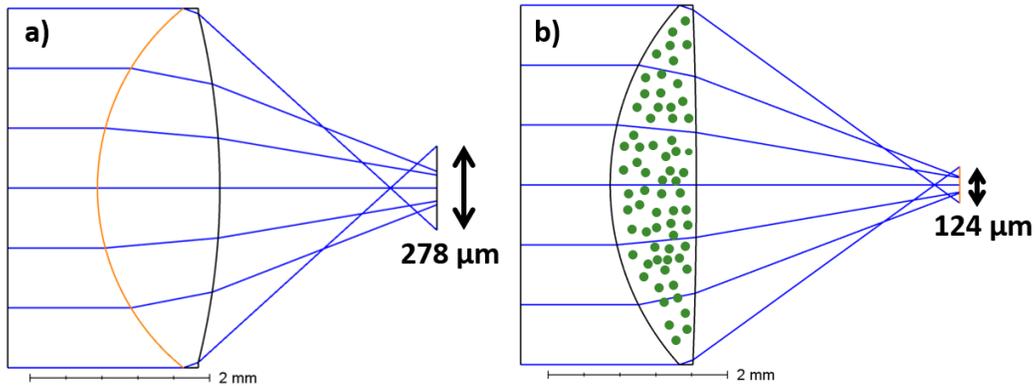

Figure 4. Design of a f=4 mm singlet with an aperture of d=4 mm without (a) and with (b) a nanocomposite material. For the design in (a) we used PMMA, whereas the nanocomposite in (b) is the $ZrO_2$/PMMA nanocomposite from figs. 4 and fig. 5 at a volume fraction of f= 35%. The use of a nanocomposite decreases the volume and increases the performance of the lens. The arrows denote the optimized spot radii (RMS) at a wavelength of 550nm.

## 5. SUMMARY AND CONCLUSION

In this paper, we have discussed that effective medium theories are powerful tools that cannot only be used to predict, and optimize the optical properties, but also provide fundamental information about how a nanocomposite has to be designed to fulfill the high transparency requirements on optical materials. This analysis has shown that different effective medium regimes have to be distinguished, and that highly transparent optical nanocomposites can only be realized in a specific range, the "unrestricted effective medium regime". Furthermore, we have shown that the use of nanocomposites as optical materials provides a new degree of freedom for optical design, and allows accessing new refractive index regimes. This provides the possibility of tailoring the optical properties of the materials to the optical system at hand, which can potentially increase its performance, while simultaneously reducing its size.

## 6. FUNDING

This project has received funding from the European Union's Horizon 2020 research and innovation programme under the Marie Sklodowska-Curie grant agreement No. 675745.

## REFERENCES

[1] Chau, J. L. H., Lin, Y.-M., Li, A.-K., Su, W.-F., Chang, K.-S., Hsu, S. L.-C., and Li, T.-L., "Transparent high refractive index nanocomposite thin films," Materials Letters 61, 2908-2910 (2007).
[2] Lee, S., Shin, H.-J., Yoon, S.-M., Yi, D. K., Choi, J.-Y., and Paik, U., "Refractive index engineering of transparent $ZrO_2$–polydimethylsiloxane nanocomposites," Journal of Materials Chemistry 18, 1751-1755 (2008).
[3] Lü, C., Cui, Z., Li, Z., Yang, B., and Shen, J., "High refractive index thin films of ZnS/polythiourethane nanocomposites," Journal of Materials Chemistry 13, 526-530 (2003).
[4] Lü, C., Cui, Z., Wang, Y., Li, Z., Guan, C., Yang, B., and Shen, J., "Preparation and characterization of ZnS–polymer nanocomposite films with high refractive index," Journal of Materials Chemistry 13, 2189-2195 (2003).
[5] Lü, C., and Yang, B., "High refractive index organic–inorganic nanocomposites: design, synthesis and application," Journal of Materials Chemistry 19, 2884-2901 (2009).
[6] Nussbaumer, R. J., Caseri, W. R., Smith, P., and Tervoort, T., "Polymer-$TiO_2$ Nanocomposites: A Route Towards Visually Transparent Broadband UV Filters and High Refractive Index Materials," Macromolecular materials and engineering 288, 44-49 (2003).
[7] Rong, M., Zhang, M., and Ruan, W., "Surface modification of nanoscale fillers for improving properties of polymer nanocomposites: a review," Materials science and technology 22, 787-796 (2006).
[8] Tao, P., Li, Y., Rungta, A., Viswanath, A., Gao, J., Benicewicz, B. C., Siegel, R. W., and Schadler, L. S., "$TiO_2$ nanocomposites with high refractive index and transparency," Journal of Materials Chemistry 21, 18623-18629 (2011).
[9] Garnweitner, G., Goldenberg, L. M., Sakhno, O. V., Antonietti, M., Niederberger, M., and Stumpe, J., "Large-Scale Synthesis of Organophilic Zirconia Nanoparticles and their Application in Organic–Inorganic Nanocomposites for Efficient Volume Holography," Small 3, 1626-1632 (2007).
[10] Luo, K., Zhou, S., and Wu, L., "High refractive index and good mechanical property UV-cured hybrid films containing zirconia nanoparticles," Thin solid films 517, 5974-5980 (2009).
[11] Chung, P. T., Yang, C. T., Wang, S. H., Chen, C. W., Chiang, A. S., and Liu, C.-Y., "$ZrO_2$/epoxy nanocomposite for LED encapsulation," Materials Chemistry and Physics 136, 868-876 (2012).
[12] Thakur, V. K., and Kessler, M. R., "Self-healing polymer nanocomposite materials: A review," Polymer 69, 369-383 (2015).
[13] Azizi Samir, M. A. S., Alloin, F., and Dufresne, A., "Review of Recent Research into Cellulosic Whiskers, Their Properties and Their Application in Nanocomposite Field," Biomacromolecules 6, 612-626 (2005).
[14] Croce, F., Appetecchi, G. B., Persi, L., and Scrosati, B., "Nanocomposite polymer electrolytes for lithium batteries," Nature 394, 456 (1998).
[15] Chatterjee, A., and Chakravorty, D., "Glass-metal nanocomposite synthesis by metal organic route," Journal of Physics D: Applied Physics 22, 1386 (1989).

[16] Gross, H., Singer, W., Totzeck, M., Blechinger, F., and Achtner, B., [Handbook of optical systems], Wiley-VCH, Berlin (2005).
[17] Hartmann, P., Jedamzik, R., Reichel, S., and Schreder, B., "Optical glass and glass ceramic historical aspects and recent developments: a Schott view," Applied Optics 49, D157-D176 (2010).
[18] Gissibl, T., Thiele, S., Herkommer, A., and Giessen, H., "Two-photon direct laser writing of ultracompact multi-lens objectives," Nature Photonics 10, 554-560 (2016).
[19] Thiele, S., Arzenbacher, K., Gissibl, T., Giessen, H., and Herkommer, A. M., "3D-printed eagle eye: Compound microlens system for foveated imaging," Science advances 3, e1602655 (2017).
[20] Rill, M. S., Plet, C., Thiel, M., Staude, I., von Freymann, G., Linden, S., and Wegener, M., "Photonic metamaterials by direct laser writing and silver chemical vapour deposition," Nature Materials 7, 543 (2008).
[21] Doyle, W., and Agarwal, A., "Optical extinction of metal spheres," JOSA 55, 305-309 (1965).
[22] Doyle, W. T., "Optical properties of a suspension of metal spheres," Physical review B 39, 9852 (1989).
[23] Bohren, C. F., and Huffman, D. R., [Absorption and scattering of light by small particles], John Wiley & Sons, New York (2008).
[24] Bohren, C. F., "Applicability of effective-medium theories to problems of scattering and absorption by nonhomogeneous atmospheric particles," Journal of the atmospheric sciences 43, 468-475 (1986).
[25] Niklasson, G. A., Granqvist, C., and Hunderi, O., "Effective medium models for the optical properties of inhomogeneous materials," Applied Optics 20, 26-30 (1981).
[26] Choy, T. C., [Effective medium theory: principles and applications], Oxford University Press, (2015).
[27] Zeng, X., Bergman, D., Hui, P., and Stroud, D., "Effective-medium theory for weakly nonlinear composites," Physical Review B 38, 10970 (1988).
[28] Malasi, A., Kalyanaraman, R., and Garcia, H., "From Mie to Fresnel through effective medium approximation with multipole contributions," Journal of Optics 16, 065001 (2014).
[29] Aspnes, D., "Local-field effects and effective-medium theory: a microscopic perspective," American Journal of Physics 50, 704-709 (1982).
[30] Skryabin, I., Radchik, A., Moses, P., and Smith, G., "The consistent application of Maxwell–Garnett effective medium theory to anisotropic composites," Applied physics letters 70, 2221-2223 (1997).
[31] Battie, Y., Resano-Garcia, A., Chaoui, N., Zhang, Y., and En Naciri, A., "Extended Maxwell-Garnett-Mie formulation applied to size dispersion of metallic nanoparticles embedded in host liquid matrix," The Journal of chemical physics 140, 044705 (2014).
[32] Markel, V. A., "Introduction to the Maxwell Garnett approximation: tutorial," JOSA A 33, 1244-1256 (2016).
[33] Mallet, P., Guérin, C.-A., and Sentenac, A., "Maxwell-Garnett mixing rule in the presence of multiple scattering: Derivation and accuracy," Physical Review B 72, 014205 (2005).
[34] Garnett, J. M., Garnett, J. M., and Heavens, O., "Optical Properties of Thin Solid Films," Phil. Trans. Roy. Soc.(London) 203, 385 (1904).
[35] Lakhtakia, A., "Size-dependent Maxwell-Garnett formula from an integral equation formalism," Optik 91, 134-137 (1992).
[36] Fearn, H., James, D. F., and Milonni, P. W., "Microscopic approach to reflection, transmission, and the Ewald–Oseen extinction theorem," American Journal of Physics 64, 986-995 (1996).
[37] Jackson, J. D., [Classical electrodynamics], John Wiley & Sons, (2012).
[38] Russakoff, G., "A derivation of the macroscopic Maxwell equations," American Journal of Physics 38, 1188-1195 (1970).
[39] Petschulat, J., Menzel, C., Chipouline, A., Rockstuhl, C., Tünnermann, A., Lederer, F., and Pertsch, T., "Multipole approach to metamaterials," Physical Review A 78, 043811 (2008).
[40] Shvets, G., and Tsukerman, I., [Plasmonics and Plasmonic Metamaterials: Analysis and Applications], World Scientific, (2012).
[41] Slavchov, R. I., and Ivanov, T. I., "Quadrupole terms in the Maxwell equations: Born energy, partial molar volume, and entropy of ions," The Journal of chemical physics 140, 074503 (2014).
[42] Werdehausen, D., Staude, I., Burger, S., Petschulat, J., Scharf, T., Pertsch, T., and Decker, M., "Design rules for customizable optical materials based on nanocomposites," ArXiv e-prints, 1808.03080 [physics.optics] (2018).
[43] Decker, M., and Staude, I., "Resonant dielectric nanostructures: a low-loss platform for functional nanophotonics," Journal of Optics 18, 103001 (2016).

[44] Wang, Y., and Herron, N., "Nanometer-sized semiconductor clusters: materials synthesis, quantum size effects, and photophysical properties," The Journal of Physical Chemistry 95, 525-532 (1991).
[45] Satoh, N., Nakashima, T., Kamikura, K., and Yamamoto, K., "Quantum size effect in $TiO_2$ nanoparticles prepared by finely controlled metal assembly on dendrimer templates," Nature nanotechnology 3, 106 (2008).
[46] Hövel, H., Fritz, S., Hilger, A., Kreibig, U., and Vollmer, M., "Width of cluster plasmon resonances: bulk dielectric functions and chemical interface damping," Physical Review B 48, 18178 (1993).
[47] Sultanova, N., Kasarova, S., and Nikolov, I., "Dispersion proper ties of optical polymers," Acta Physica Polonica-Series A General Physics 116, 585 (2009).